# Identity Management through a global Discovery System based on Decentralized Identities

Konstantinos Lampropoulos
*Department of Electrical & Computer Engineering*
*University of Patras*
Patras, Greece
klamprop@ece.upatras.gr

Nikos Kyriakoulis
*Department of Electrical & Computer Engineering*
*University of Patras*
Patras, Greece
kyriakoulis@ece.upatras.gr

Spyros Denazis
*Department of Electrical & Computer Engineering*
*University of Patras*
Patras, Greece
sdena@upatras.gr

*Abstract*— Digital identities today continue to be a company resource instead of belonging to the actual person they represent. At the same time, the digitalization of everyday services intensifies the "Identity Management problem" and leads to a constant increase of users' online identities and identity related data. This paper presents DIMANDS2, a framework capable of organizing identity data that allows service providers and identity issuers securely exchange identity related information in a privacy-enabled manner while the user maintains full control over any activity related to his/her identity data. The framework is "format-agnostic" and can accommodate any type of identifier (existing or new), without requiring from existing services and providers to implement and adopt another new global identifier.

## I. INTRODUCTION

The different definitions of "identity" and "identity management (IdM)", reveal an underlying issue about these terms in the digital world. The first one defines identity as "*the distinguishing character or personality of an individual*" [1][2] while the second describes identity management as "*a framework of policies and technologies for ensuring that the right users (in an enterprise) have the appropriate access to technology resources* [3][4]. Thus, even though identities are supposed to distinguish (and belong to) individuals (users), their creation and management (in the digital world) is not controlled by them, but by service providers inside closed contexts. The "Identity Management problem" is further aggravated by the constant growth of digital services and domains which introduce diverse administration models, rigid processes for provisioning and deprovisioning identities, user password fatigue etc. [5] [6]. For many years now, efforts to address the IdM problem, usually examined only some of its aspects inside very specific contexts with highly diverse requirements. The result was numerous different IdM solutions with serious interoperability issues among emerging IdM islands. Attempts to integrate these islands (also called Identity Federations) recursively introduce new identities, procedures and eventually new islands (in a fractal-like fashion that repeated itself).

The latest years, research on identity management has taken advantage of various ICT innovations like blockchains, decentralized file systems, Decentralized Identities - DIDs [7] etc. to create new solutions for self-sovereign identities. But even though these systems do offer considerable advances, a closer analysis (as presented by the authors of [8]) reveals that the advertised decentralization is in most cases a "reshape of the role of centralization and intermediaries" to different hosts and organizations. Another important issue that is often not tackled at all, is user experience and behavior. Designers seem to assume that the average user can do complicated actions (e.g., effective key management) or can understand the implications of putting identity data or attributes in a DLT[8]. For example, DIDs give users full control over their identities when reports indicate that internet users always take the easy path and will directly submit whatever information is asked from them (e.g., only 0.33% disable one or more cookies categories [9]).

The complexity of the "Identity problem" cannot be addressed with the integration of identities since new services and systems will always be created using their own formats and protocols. We argue that a viable solution is feasible through the dynamic association and discovery of identity data which exist (now and in the future) and operate autonomously inside their own contexts and under their own requirements. This paper presents DIMANDS2, (Digital Identity MAnagement N' Discovery System) which is a system based on the work initially published in [10]. DIMANDS2 improves the previous system (DIMANDS) by introducing a completely new architecture based on existing technologies and frameworks (Decentralized Identifiers (DIDs), Verifiable Credentials (VCs) [11]) to provide a more secure and easy to implement approach. This approach also introduces the use of a mobile application that allows the user to overview all operations on his/her identity data, thus significantly improving the usability of the system and making it more attractive to inexperienced users.

The main contributions of DIMANDS2 are: i) it is a system designed to organize users' identity information which until now reside scattered across multiple isolated contexts; ii) it allows service providers to exchange/share identity related information in a privacy-enabled manner emphasizing on minimal disclosure of information; iii) it provides to the user full control over any all activities performed on his/her data; iv) it is "format-agnostic" in the sense that it can accommodate any type of identifier (existing or new); v) like its predecessor, DIMANDS2 does not store any user information and achieves high levels of privacy and security offering resistance against profiling attacks and malicious acts, coming from any internal or external entities (see section V).

The rest of the paper is organized as follows. Section 2 presents the current state of the art and section 3 our proposed architecture. In section 4 we present how the system can support cross domain identity services and section 5 provides a security and privacy analysis. Finally, section 6 concludes the paper discussing open issues and future work.

The research leading to these results has received funding from Horizon 2020 Research & Innovation Programme CONCORDIA under grant agreement No 830927

## II. STATE OF THE ART

Identity management has been an intensively researched topic for many years now and there are multiple systems and research solutions that have been proposed so far.

On Web standards and specifications, the OASIS XRI Data Interchange (XDI) [12] is an effort to define a generalized, extensible service for sharing, linking, and synchronizing data over the Internet using XML documents and XRIs (Extensible Resource Identifiers). Blockcerts [13] is an open standard for building apps that issue and verify blockchain-based official records like certificates for civil records, academic credentials, professional licenses, etc. W3C has release a specification for a) Decentralized identifiers (DIDs) which are a new type of identifier that enables verifiable, decentralized digital identity and b) Verifiable Credentials Data Model [11], which are the electronic equivalent of the physical credentials that we all possess today, like passports, driving licenses, etc. The Identifiers & Discovery Working Group – DIF [14] is working for the development of protocols and systems that enable creation, resolution, and discovery of decentralized identifiers and names across underlying decentralized systems, like blockchains and distributed ledgers.

In the industry world, there are multiple cases of companies which adopt and extend many of the aforementioned standards and specifications. Some examples include Microsoft with Microsoft Identity and ION network [15], IBM with IBM identity [16], Sovrin [17], uPort [18], ShoCard and Ping Identity [19], etc. At the same time the eIDAS [20], the European Self-Sovereign Identity Framework (ESSIF) [21], e-Residency and e-identity [22], along with the California Consumer Privacy Act (CCPA) [23] and the EU General Data Protection Regulation (GDPR) [24] are some the solutions that the public sector is adopting for supporting citizens with the management of identities and personal information. These solutions (from the industry and public sector) can offer considerable contributions and high levels of security; however, such approaches shift the centralization to different actors and present distributed systems that are still owned by a central entity/authority. Furthermore, they focus on very specific aspects of the IdM problem and are usually operational inside very specific contexts (e.g., public administration services etc.). Contrary to these solutions, DIMANDS2 creates a format agnostic IdM discovery and management system that goes beyond the boundaries of specific contexts.

In the academic and research fora, publications generally describe systems that use blockchains as the underlying technology to store users' identities, or representations of them. To be able to balance the increased levels of privacy requirements with the transparency features of distributed ledgers these initiatives either introduce off-chain components (e.g., Blockstack Global Naming System [25]) or use permissioned blockchains (e.g., Anonymous Identities for Permissioned Blockchains [26]). Other research papers on identity management and blockchain technologies include BPDIMS [27], AttriChain [28], LifeID[29], Sora identity[30] reclaimID[31], etc. Such works introduce diverse approaches on the use of blockchain for self-sovereign identity and build upon DIDs, VCs and blockchains to support IdM in special contexts and cases. Contrary to proposed frameworks that assume that users can manage all their identity data (DIDs) alone, DIMANDS2 is built to help users and hide the complexity of operations. With DIMANDS2 identity data remain stored inside the secure contexts of identity issuers, and identity requesters (service provider) can retrieve them through transparent processes that can be approved or rejected by the user.

Finally, the DNS-IdM [32] system discusses discovery aspects inside an IdM system that uses both a permissioned and a permissionless blockchain. It introduces the "DNS contract" to locate (discover) the address of a contract validator for a specific attribute. In DIMANDS2, the discovery aspect focuses on the identification (discovery) of a user through any given identity. We argue that such a system has not yet been proposed, contrary to the many other DNS-like systems that already exist for service discovery.

## III. DIMANDS2 ARCHITECTURE

As mentioned above, DIMANDS2 makes use of Decentralized Identifiers and Verified Credentials specifications to introduce a global identity discovery and association system. We need to clarify here that DIMANDS2 adopts the basic models of DIDs and VCs and makes alterations (modifications) to support its desired functionality. These alterations are specific rules that apply to some of these specifications' fields. The reason we adopted the DIDs and VCs instead of introducing our own models, is because we argue that our changes are minor (and not too complex), thus if DIMANDS2 is to be widely accepted by the community, its components can be easily implemented as extension of established W3C specifications.

### A. Architecture

The first component of DIMANDS2 architecture is the D2-Hub, a node that resembles the functionality of Identity Hubs [33] in the sense that users may have their own accounts to store and control their own identity related information. This component might be in the future an extension of the current form of Identity Hubs, however in this paper we do not consider them to be. D2-Hubs do not store any actual user identities and can be operated by any kind of organization (e.g., universities, network providers, large enterprises etc.). To be able to fully trust this component we propose that D2-Hubs must be validated by an overviewing authority (e.g., ICANN). An D2-Hub must be publicly accessible under a discoverable ID (domain name, DOI etc.). The type of this ID does not affect the overall functionality of the system, thus for this version of the system we assume that D2-Hub IDs are URLs.

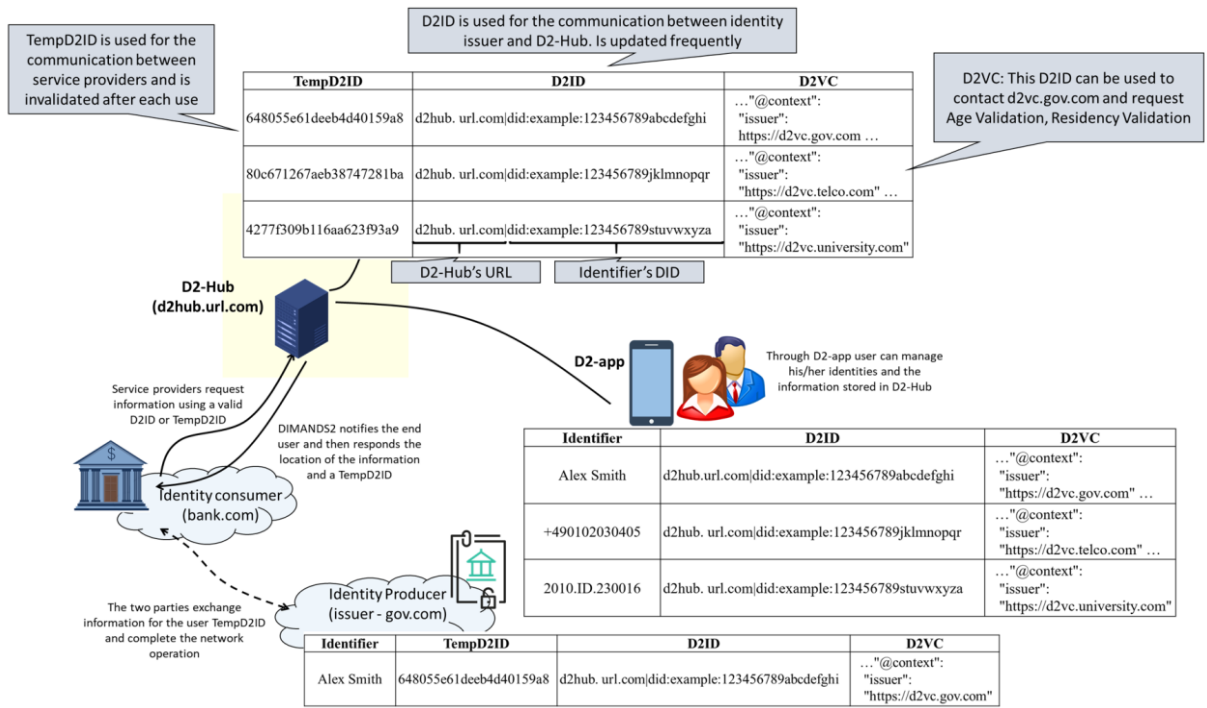

In DIMANDS2 users may select the D2-Hub they prefer and create an account. Each account has a database which holds information related to its owner identities. This database has three fields (Fig. 1). The "TempD2ID", "D2ID" and "D2VC". The D2ID is a unique representation of a specific identity issued by a service provider. A D2ID can represent an identifier of any type, format etc. that exist in any kind of context. It is formed as the concatenation of the D2-Hub's discoverable ID (URL) and a DID which is created by the service provider that issued this identity. Using the URL (first part) of a D2ID a request can be sent to the appropriate D2-Hub. Then, by using the second part of the D2ID (the DID), D2-Hub can identify the user that the request is related to.

Each D2ID -and thus associated identifier- can be used to perform identity operations regarding a user. To enable this functionality, we introduce our own kind of credentials document called D2VCs. A D2VC is based on a "modified" W3C Verifiable Credentials specification which includes only the "type" and "issuer" fields. The "issuer" field stores the URL of the service provider that created the original identifier and the "type" field describes one or more capabilities that this D2ID can support. Such functionality can be "Authentication", "Age Validation", "Proof of Nationality" etc. This field does NOT contain any actual data (D2VC does not have a "claim" field) and cannot reveal any kind of user's identity information. For example, the D2VC type "Age Validation" for a specific D2ID (and corresponding identity) refers to the ability of the service provider that produced this identity to validate a user's age and does not contain any information about user's actual age like a Verifiable Credential would.

Finally, the TempD2ID value is a completely random identifier (must be a unique value inside a single D2-Hub) and it is used to avoid sharing the D2ID outside DIMANDS2. Our goal is to use the D2ID only for the communication between the service provider that issued the identity associated with it and the D2-Hub. In all other cases (e.g., communication between two service providers) the involved parties must use a temporary identifier (TempD2ID) that is invalidated after each use. Details about the usage of D2ID, TempD2ID and D2VCs is demonstrated in the next section. At this point we must clarify that the D2ID and TempD2ID values are not introduced as new global identifiers. They exist only inside DIMANDS2 and are not expected to replace real identities or affect existing protocols in today's (or future) services and contexts.

Users may securely connect to their D2-Hub account to control their identities using their phone and a mobile app called D2app. This application is a key component of the architecture, since it is the only point where all information (real identities, D2IDs, credentials, rules etc.) are stored and linked together. D2app provides the interface (UI) through which users are informed (by D2-Hub) every time a service provider requests access to their identity information (information that exists inside an identity producer-issuer). There, users can control whether they will allow, reject or even choose which identity producer may provide the requested data. D2app also provides functionality for registration, deletion or modification of new identities, creation of policies and rules etc. It stores all its data locally in an encrypted form (e.g., app requires extra pin to unlock) and allows for data extraction (also in encrypted form) for recovery purposes (e.g., loss or change of a device). The overall architecture of DIMANDS2 is presented in Fig.1.

### B. Registering a new Identity to DIMANDS2

The process of registering a new username (a new identifier created by a service provider like e.g., a new email) to DIMANDS2 is done through D2app. Using the option "Register new identity" (Fig.2) the user is requested to insert the service provider URL that issued the corresponding username. Then he/she is redirected to the service provider to login and prove that he/she is the legitimate owner of the username he/she wants to register to DIMANDS2. The credentials for this login can be given to him/her at the time of creation of this new identity. After the user is validated, the service provider creates a new DID, and D2VC data with all the capabilities supported by this identity (e.g., ability to validate user's nationality, age etc.).

Once the DID and D2VC data are created, the service provider sends them back to D2app. Then the D2app adds the D2-Hub ID (URL of D2-hub) to the DID value, to produce the D2ID and stores it locally. New values (D2ID and D2VC) are transmitted to D2-Hub and back to the service provider to be added in the user's profile. The last step is for the service provider to create a TempD2ID and transmit it to D2-Hub to be associated with the corresponding D2ID. The process of registering a new identity to DIMANDS2 must be done only once. After it is completed, the service provider is responsible for frequently updating the TempD2ID to ensure that the framework's high levels of privacy are maintained. At the end of this new identity registration, the service provider's database contains the values Identifier/ TempD2ID/D2ID/D2VC, the D2-Hub database contains the values TempD2ID/D2ID/D2VC, and user's application (local storage) contains the values identifier/D2ID/D2VC (Fig. 2).

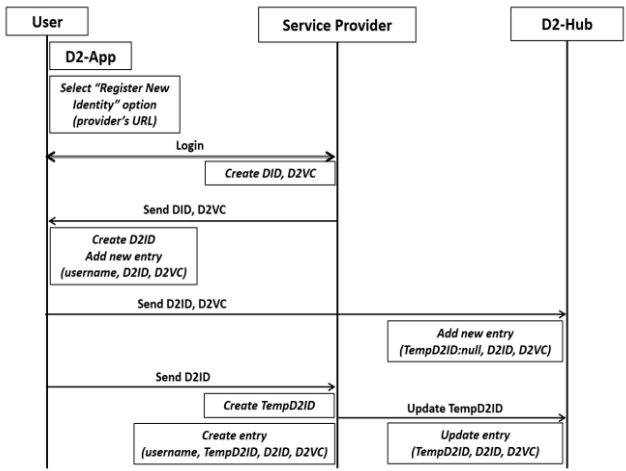

Fig. 2: Registering a new identity to DIMANDS2

### C. Service providers and DIMANDS2

The main goal of DIMANDS2 is to avoid introducing new global IDs which must be adopted by existing services and protocols. Thus, all new data types used for DIMANDS2 values (D2ID, TempD2ID etc.) are only meaningful inside the context of DIMANDS2 and do not impose the need for changing existing services, networks, protocols etc., to use them. For a service provider, to be able to join DIMANDS2, it only needs to a) add in its user databases the required fields to associate existing users' identifiers (usernames) with the TempD2ID, D2ID and D2VC values and b) implement the APIs for exchanging data inside the DIMANDS2 context.

## IV. DIMANDS2 FUNCTIONALITY

In this chapter we present how DIMANDS2 can be used to address identity-related challenges of the KYC (Know Your Customer) process inside the banking sector. With this scenario we will depict how DIMANDS2 can help a service provider discover any kind of identity related information it needs to provide a service and acquire this information from trusted sources. At the same time the user maintains full control over the process without the need to be involved in complicated tasks or take decisions about selecting the correct amount of information that is required to share (which inexperienced users usually find difficult to manage).

### A. Know your Customer (KYC)

In the banking sector, KYC is a mandatory process and involves the verification of a client when opening a bank account, and periodically for as long as the account is active. Users must submit either physical or scanned documents (ID, recent utility bill etc.) which makes it difficult and time consuming for the banks to verify their validity. DIDs are tackling this issue to a specific point, but inexperienced users still find difficult to use these new technologies and most of all select the correct amount of information that is actually needed.

Fig. 3 presents how DIMANDS2 may facilitate and further secure this process. With DIMANDS2, when a user visits a bank to open a new account, the bank will not request any information directly from him/her. Instead, it will only ask an identity that is registered (and resolvable) through DIMANDS2. Such an identity can be a simple email. Once the user provides this email, then the bank will send a DIMANDS2 request to the "email.com" provider asking for the D2-Hub that the user@email.com has a registered account and a TempD2ID (The "email.com" provider can retrieve the D2-Hub URL from the first part of the identifier's D2ID).

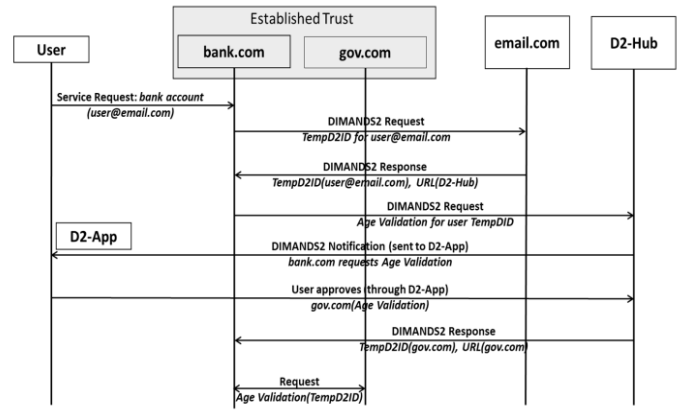

Fig. 3: Age Validation for KYC

With these values, the bank can now directly contact the D2-Hub, asking information about the user (using the TempD2ID value). In this scenario, this request involves the verification of user's age (Age Verification) and address (Address Verification). With the TempD2ID value, the D2-Hub can find the user that this request is targeted to and send an alert to his/her phone (D2app) informing him/her that provider "bank.com" wants to validate his/her age and address. Along with this alert, the D2app can present to the user a list of providers (where he/she already has an account) that can answer this request. The user selects the two providers that wishes to verify his/her age (gov.com) and address (telco.com) and informs the D2-Hub about his/her selection. D2-Hub retrieves from the user's account the TempD2IDs for each provider and sends them to the bank. Now the bank can directly contact each provider separately and retrieve information to verify the user.

Notes about this scenario:

*1)* The TempD2ID value is used to locate a user inside a service provider without revealing his actual username, DID or D2ID. A user can only be identified by a TempD2ID once, and after a TempD2ID is used in any kind or transaction, the service provider is required to update it (create a new one) in his database and the D2-Hub. Thus, the user identity inside the service provider is always protected, and when someone wants to locate him/her again inside the same provider, the user will be notified again to allow it or not.

*2)* After successful validation of the customer, and the opening of the account, the user can register his bank identity (username) in DIMANDS2. Then the bank will have a D2ID for this user that can use for any future references or identity related activities (e.g., periodic validation of user). D2ID values are only used between the service provider that owns the corresponding identity and D2-Hub. Thus we assume that their frequent update is not as necessary as the update of TempD2ID (which can be shared between multiple service providers).

*3)* One of the innovations of DIMANDS2 is that when the bank receives the response containing the providers that can verify the new customer (in our case gov.com and telco.com), it can decide whether it trusts them or not. As long as the bank trusts the two providers, then the amount of information to be shared can be absolutely minimun. For example in our case the "gov.com" provider can simply acknowledge that the user is "over 18", without revealing his/her actual age (zero-knowledge proof). The actual method of validating user's credential is not part of DIMANDS2 architecture. We argue that the final message exchange should be an open process, where the different parties that trust each other may independently select the method/protocol/model based on which they can complete the identity operation. DIMANDS2 contribution stops at the point where these two parties can securely identify a user without the need to share his/her actual identity.

*4)* In our scenario, the bank directly contacts each provider to retrieve the necessary information. This communication however allows the two providers (gov.com, telco.com) learn that their user will create a new account at a specific service provider. There might be cases where the user many not want the validation provider to learn about his new registration (e.g., when a user must properly validate his age in a healtcare provider, using government credentials). DIMANDS2 can also cover this scenario through a different message exchange process. In particular, when the new service provider (e.g. healthcare) selects the organization (e.g. gov.com) that trusts to validate the age of its new user, it does not directly contact it. Instead it replies back to D2-Hub its selection and a public key. The D2-Hub forwards the request and the public key to gov.com which in turn replies back with a validation response. This response is signed to ensure that it is issued by the specific issuer (gov.com) and encrypted (with the requester's public key) to ensure that D2-Hub is not able to see or tamper the message. Finally D2-Hub forwards it the encrypted service message to the service provider (healthcare) to complete the validation procedure.

## V. SECURITY, PRIVACY AND USABILITY

In this paper we describe a discovery mechanism that can connect the various identities that a user has across different contexts. The association of users' identifiers has been proposed for many years now but has not been implemented due to security and privacy concerns. DIMANDS2 is designed to provide high levels of security, privacy (security by design) and usability. Below we present an overall analysis about various security, privacy and usability aspects of the proposed system and components.

*1) D2-Hub*: The first important aspect we need discuss is why this component is necessary and why an online hub is needed, when D2app already offers a point where all user's identities converge. In our architecture D2-Hub is necessary to provide a level of separation between the end user and the various service providers. It also allows service providers to initiate requests when looking for information about a user, and at the same time the user be informed about these requests through one active connection to the D2-Hub. Without the use of D2-Hub (e.g., using a mobile wallets of identities), communication could only be initiated by the user (since his phone will almost always reside behind a NAT). D2-Hub component is designed to offer high levels of security and privacy. It stores only representations of users' identities (D2IDs and TempD2IDs) and not the actual identities. The only visible information stored in D2-Hub is the domains that a user has accounts and their capabilities. Thus we argue that an attack targeting to steal D2-Hub's databases cannot easily result in profiling. Even if such an attack is successful users have the ability to completely renew all D2IDs and TempD2IDs (through automated requests to their service providers), and quickly invalidate all leaked information. One special case involves an attack performed by a malicious service provider that can identify one (or many) of its users from a D2ID he already possesses. In this case the service provider will be able to learn other service providers that its users have accounts in (only the providers and not the actual usernames). This case can also be avoided by having D2-Hubs encrypt their databases. As mentioned above, D2-Hub operator is validated and overviewed by an authority (e.g., ICANN) to ensure that it is a trusted organization that does not try to act maliciously against its users. The architecture of DIMANDS2 also provides extra security by allowing users to select the D2-Hub provider they trust the most. Furthermore, it supports the invalidation and migration of accounts between D2-Hubs. Finally, considering the case of a D2-Hub providing wrong links to identity requesters (as described in Section II), this is more of a functionality issue and not an security threat. Identity requesters are expected to contact they issuers that already know and trust. Thus any wrong links sent to them will only result in failure to complete the operation and not to any kind of data leakage. Finally about the security when an identity requester and an identity issuer do not directly communicate (but use D2-Hub as described in note 4 in the KYC scenario) we argue that it is ensured since the identity issuer signs and encrypts its response.

*2) D2App:* The D2App application is one of the most essential parts in DIMANDS2 architecture thus any security flaw may cause serious data exposure. For this component we argue that it should be an open source project to allow the community always check the code for security vulnerabilities. Considering the data stored locally, they must be encrypted and should only be accessible after user authentication. Copies of these data (to be used for recovery or device change) should only be made in encrypted form. D2App stores the identities of a user and not any kind of identity related information (credentials) that can prove claims for him/her (this information will always reside inside the issuer's domain). Thus we consider it to be more safe, compared to other approaches (identity wallets) that store all the identity data in users' devices. Also as also mentioned above, by connecting the service providers with the identity providers (issuers) we a) remove from the inexperienced users the burden to make decisions or use technologies that usually find hard to manage and b) allow information exchange in a way that is transparent, minimises data exposure (zero knowledge proof) and can be overviewed by users and -if necessary- authorities.

## VI. Conclusions and future work

In this paper we presented DIMANDS2, an identity association and discovery system capable of organizing identity data that reside scattered across multiple isolated contexts. We described its architecture, basic components and advances compared to its earlier versions and demonstrated through a clear scenario how it can address significant identity-related issues of an existing widely used banking service.

The system is currently implemented in demo version and will be tested inside a European funded project that aims to build a cybersecurity competence center for EU to be further evaluated and evolved. Future work will focus on the implementation of an alpha version that can be publicly shared as well as tested with actual users of different security background. Another aspect that will also be investigated will be the adoption (or design) of one or more protocols that will facilitate the final identity data exchange between identity requesters (service providers) and identity issuers. As mentioned above, currently this process is outside the scope of this work leaving the communicating parties agree on the method that better suits their service. However, further research is required to at least identify (or design) basic methods/protocols to facilitate this exchange and offer a complete solution for identity management.